\documentclass[showpacs,twocolumn,aps,pre]{revtex4}
\usepackage{amssymb}

\usepackage{graphicx}
\usepackage{dcolumn}
\usepackage{bm}

\def\be{\begin{equation}}
\def\ee{\end{equation}}
\def\bea{\begin{eqnarray}}
\def\eea{\end{eqnarray}}
\def\ba{\begin{array}}
\def\ea{\end{array}}
\def\a{\alpha}
\def\b{\beta}
\def\c{\gamma}

\def\d{\delta}

\def\0{$\Gamma_0$}
\def\o{\omega}

\def\r{\rho}
\def\t{\theta}

\begin{document}

\title{Duality between quantum and classical dynamics for integrable billiards}
\author{W. T. Lu}
\email{w.lu@neu.edu}
\author{Weiqiao Zeng} 
\author{S. Sridhar}
\email{s.sridhar@neu.edu}
\affiliation{Department of Physics and Electronic Materials Research
Institute, Northeastern University, Boston, Massachusetts 02115}

\date{\today }

\begin{abstract}
We establish a duality between the quantum wave vector spectrum and the
eigenmodes of the classical Liouvillian dynamics for integrable billiards. 
Signatures of the classical eigenmodes appear as peaks in the correlation
function of the quantum wave vector spectrum. A semiclassical derivation and
numerical calculations are presented in support of the results. These
classical eigenmodes can be observed in physical experiments through the
auto-correlation of the transmission coefficient of waves in quantum billiards.
Exact classical trace formulas of the resolvent are derived for
the rectangle, equilateral triangle, and circle billiards.
We also establish a correspondence between the classical periodic orbit length  
spectrum and the quantum spectrum for integrable polygonal billiards.
\end{abstract}

\pacs{02.30.Ik, 03.65.-w, 05.20.Gg, 03.65.Sq}
\maketitle

\section{Introduction}

Two sets of spectra can be associated with any dynamical system. The quantum
spectrum, represented by the eigenvalues of the Schr\"{o}dinger equation, is
a set of real numbers for a closed system, while for an open system, the
eigenvalues are complex and are called resonances. The classical dynamics
can be described in terms of the spectrum of eigenvalues of the Liouville
operator. For hyperbolic systems, the classical spectrum is comprised of the
so-called Ruelle-Pollicott resonances \cite{Ruelle,Pollicott} which
determine the time evolution and relaxation of classical correlations.

Since the birth of quantum mechanics the correspondence between the
classical and quantum properties has been examined from different
perspectives. While the classical dynamics in the phase space is governed by
the Liouville equation, the corresponding quantum dynamics in phase space is
governed by the Moyal equation of the Wigner function. In the classical
limit, the higher order terms of $\hslash $ in the Moyal equation vanish,
and one retrieves the classical Liouvillian dynamics. This correspondence in
the phase space properties has been studied by Brumer and collaborators \cite
{Brumer97}. In this paper, we obtain three principal results that establish a duality 
between quantum and classical dynamics in integrable billiards. 

For billiard systems, it is natural to work in wave vector space \cite{Miller,Cohen}. We
find that the two-level correlation of billiard systems in wave vector $k$
space is invariant with respect to the correlation interval. Peaks are
observed in the quantum spectral correlations which are shown to be centered
at the classical eigenmodes of the Liouvillian dynamics. Thus we establish a
duality between the quantum wave vector spectrum $\{k_{n}\}$ and the
classical eigenmodes $\{\gamma _{n}\}$ for integrable closed billiard
systems, which can be expressed as 
\begin{equation}
C[f(\{k_{n}\})]=g(\{\gamma _{n}\}).
\end{equation}
Here $f(\{k_{n}\})$ and $g(\{\gamma _{n}\})$ are certain spectral functions
with $C[f]$ an appropriately defined quantum correlation. The quantum wave
vector spectral correlation directly leads to the classical eigenmodes. The
results are numerically demonstrated for example systems including the
two-dimensional rectangle, equilateral, and the circle billiards. A
semiclassical derivation is provided supporting the results. Generalization
to higher dimensional integrable systems is straightforward.

It is well known that the quantum spectrum can be calculated from classical 
periodic orbits (POs) through the Gutzwiller trace formula. 
Here we obtain an exact inverse result, viz., that the classical PO
length spectrum can also be expressed in terms of the quantum wave vector spectrum, for 
integrable polygonal billiards. Finally, we also obtain exact classical 
trace formulas for integrable billiards. 

In Sec. II, the quantum correlation is defined and the billiard systems
under consideration are introduced. The corresponding classical dynamics are
discussed, the classical eigenmodes and exact classical trace formula
are obtained in Sec. III. A semiclassical derivation
of the quantum correlation is provided in Sec. IV. Detailed numerical
calculations supporting the main results are presented in Sec. V. Exact relation
between the classical POs spectrum and the quantum
spectrum is established for integrable billiards in Sec. VI. Finally, 
Sec. VII presents a summary and some remarks on the results.

\section{Quantum eigenvalues and correlations}

In this paper, we consider the wave vector density of states (DOS) as our
quantum spectral function which is a sum of $\delta $-functions $\rho
(k)=\sum_{n}\delta (k-k_{n})$. In order to study the quantum spectral
correlation, we use a Lorentzian-smoothed DOS 
\begin{equation}
\rho _{\epsilon }(k)=\frac{1}{\pi }\sum\limits_{n}\frac{\epsilon }{%
(k-k_{n})^{2}+\epsilon ^{2}}.  \label{dos-1}
\end{equation}
Here $\epsilon $ is a small width. In the limit $\epsilon \rightarrow 0$,
one gets the stick spectrum. The continuous part of the DOS for
two-dimensional closed billiard systems is obtained from the Weyl law as $%
\left\langle \rho (k)\right\rangle =(A/2\pi )k\pm L/4\pi $ with $A$ and $L$
the area and perimeter, respectively. The plus sign is for Neumann boundary
condition (NBC) while minus sign for Dirichlet boundary condition (DBC). The
fluctuation part of DOS is $\delta \rho _{\epsilon }(k)=\rho _{\epsilon
}(k)-\left\langle \rho (k)\right\rangle $. Define the following correlation 
\begin{equation}
C_{\rho _{\epsilon }}(s)\equiv \left\langle \delta \rho _{\epsilon
}(k)\delta \rho _{\epsilon }(k+s)\right\rangle _{k}.  \label{cor-1}
\end{equation}
Here $\delta \rho _{\epsilon }(k)$ is with a window $[K_{0}-\Delta
,K_{0}+\Delta ]$ with certain $K_{0}$ and $\Delta $ such that $\Delta \gg
\delta $, the mean level spacing. The range of average over $k$ is from $%
-\Delta -\min (0,s)$ to $\Delta -\max (0,s)$. The range of $s$ can be safely
put as $-\Delta <s<\Delta $. The average over $k$ is defined as $\langle
f(k)\rangle =(b-a)^{-1}\int_{a}^{b}f(k)dk$. 
Cross-correlation can be defined similarly for any two intervals 
$[K_1-\Delta, K_1+\Delta]$ and $[K_2-\Delta, K_2+\Delta]$ with 
arbitrary $K_1$ and $K_2$ \cite{Lu04}.

We remark that the correlation $C_{\rho _{\epsilon }}(s)$ in Eq. (\ref{cor-1}%
) is essentially the two-level correlation $R_{2}(s)$. The difference is
that here, there is no unfolding and no rescaling by the mean level spacing
of the energy levels. \emph{For billiard systems, the correlation }$C_{\rho
_{\epsilon }}(s)$\emph{\ in Eq. (\ref{cor-1}) is invariant with the choice
of }$K_{0}$\emph{\ and }$\Delta $\emph{\ as long as one has enough quantum
eigenvalues in the interval}. For the Riemann zeros \cite
{Berry307,Bohigas01,Leboeuf01,Leboeuf04,Lu04}, since the mean density is $%
\left\langle \rho (k)\right\rangle =(1/2\pi )\ln (k/2\pi )$ which is very
flat and practically a constant for large $k$, there is no real difference
between unfolding and not-unfolding. But for billiard systems, not-unfolding
is essential to uncover the structures in the quantum spectral correlations.

Here we focus on the quantum spectra of three integrable systems, the rectangle, 
equilateral triangle, and circle billiards.

\emph{Rectangle billiard} For rectangle billiard with sides $a$ and $b$, the
eigen wave vector is 
\begin{equation}
k_{mn}=\pi \sqrt{(m/a)^{2}+(n/b)^{2}}.  \label{eig-rect}
\end{equation}
Here for DBC $m,n\geq 1$ and $m,n\geq 0$ for NBC.

\emph{Equilateral triangle billiard} The equilateral triangle billiard is
integrable only if all the sides have the same boundary condition. The
eigenvalues are given by 
\begin{equation}
k_{mn}=(4\pi /3a)\sqrt{m^{2}+n^{2}-mn}  \label{eig-tri}
\end{equation}
with $a$ the side length. For the equilateral triangle with DBC, $m\geq
2n\geq 2$ while with NBC, $m\geq 2n\geq 0$.

\emph{Circle billiard }The circle billiard is integrable. The spectrum $%
\{k_{mn}\}$ is given by 
\begin{eqnarray}
J_{m}(k_{mn}a) &=&0,\text{ DBC},  \nonumber \\
J_{m}^{\prime }(k_{mn}a) &=&0,\text{ NBC}.
\end{eqnarray}
Here $a$ is the radius of the circle and $J_{m}(x)$ is the Bessel function
of the first kind.

\section{Classical eigenmodes and trace formula in Liouvillian dynamics}

We first discuss the classical eigenmodes of the Liouvillian dynamics. For
Hamiltonian systems, the phase space density distribution $\varrho (q_i,p_i,t)$ is
governed by the Liouville equation 
\begin{equation}
\frac{\partial}{\partial t}\varrho (q_i,p_i,t)={\hat L}\varrho (q_i,p_i,t)
\end{equation}
where ${\hat L}\equiv\{H, \ \}$ is the Liouville operator. 
Since we are interested in the spectrum of ${\hat L}$ instead of the 
detailed solutions of the equations of motion, it is better to switch 
to the action-angle variables which are related to
$q_i$ and $p_i$ through a canonical transformation \cite{Goldstein}.
In the action-angle
variables, the Hamiltonian is a function of actions 
$I_{i}=(2\pi)^{-1}\oint p_idq_i$, $H=H(I_{i})$. 
For fixed $I_i$, the phase space is a N-dimensional torus 
with area $(2\pi)^N$ for a N-dimensional integrable system.
The equations of motion for the angle variables $\vartheta_i$ 
are $d\vartheta_i/dt=\o_i$ with $\omega _{i}\equiv\partial H/\partial I_{i}$,
which give solutions as linear functions of time.
These frequencies have dependence on action variables $I_{i}$ and are in general
continuous. They are often associated with certain trajectories. 
Since the Hamiltonian is cyclic in $\vartheta_i$, 
the Liouville operator ${\hat L}$ assumes a much simpler form in terms 
of action-angle variables
\be
{\hat L}=-\sum_{i}{\partial H\over \partial I_i}{\partial \over \partial \vartheta_i}
=-\sum_{j}\o_j{\partial \over \partial \vartheta_j}.
\ee
The eigenvalues of operators $i{\partial \over \partial \vartheta_j}$ all take
integer values for motions on the torus. Thus $\omega _{i}$ are the 
primitive eigen frequencies of the Liouville operator ${\hat L}$ on the 
$N$-dimensional torus.

Except for one-dimensional systems, these frequencies $\o_i$ are not 
the eigenmodes of the Liouville operator on the energy surface which has $2N-1$ dimensions. 
In order to obtain the classical eigenmodes of the system, an ensemble average must be 
performed over the whole energy surface.
A good way is to look at the trace of the classical evolution operator 
\begin{equation}
\mathrm{tr}\ e^{{\hat L}t}={1\over \mu}\sum_{\mathbf{n}}\int d\mu
\exp (i\mathbf{n}\cdot \mathbf{\omega }t).  \label{trace-freq}
\end{equation}
Here $d\mu$ is the Liouville measure on the energy 
surface in phase space \cite{ODA}. 
For two-dimensional integrable systems, one has
\be
d\mu=4\pi^2 \d(E-H(I_1,I_2))dI_1dI_2.
\ee
Here the factor $4\pi^2$ is from the integration over the two angle variables 
since any integrand we will consider for an integrable system is 
independent of the angle variables.
The total measure is $\mu=4\pi^2\int\int dI_1dI_2\d(E-H)=dV/dE$ 
with $V$ the phase space volume.
The Laplace transform will give the trace of the resolvent 
\begin{equation}
g(z)\equiv \mathrm{tr}(z-{\hat L})^{-1}=\int_{0}^{\infty }\mathrm{tr}\
e^{{\hat L}t}e^{-zl}dl.
\end{equation}
Here for billiards, the particle speed $v$ is set to be unity, 
thus time $t$ is identified with length $l$ throughout the paper. 
The classical eigenmodes will show up as singular points in the trace of the
resolvent. For hyperbolic systems, the trace of the resolvent has simple
poles which are called the Ruelle-Pollicott resonances $g(z)=\sum_{n}
d_{n}/(z-\gamma _{n})$ with $d_{n}$ the degeneracy. For generic systems, the
classical trace $g(z)$ will have various kinds of singularities which
can be associated with the classical eigenmodes.

\emph{Rectangle billiard} For a particle with mass ${\mathfrak m}$ 
moving inside a rectangle billiard, the two actions are
$I_x=pa\cos\a/\pi$ and $I_y=pb\sin\a/\pi$ with $p=\sqrt{2{\mathfrak m}E}$ and $\alpha $ the
angle of the velocity with respect to the $x$-axis. One thus gets $\omega _{x}=(\pi
v/a)\cos \alpha $ and $\omega _{y}=(\pi v/b)\sin \alpha $ 
with $v$ the particle speed in the billiard. Since the Jacobian determinant 
$|\partial(I_x,I_y)/\partial(p,\a)|$ is $pab/\pi^2$, so one has
$d\mu=4ab\d(E-p^2/2{\mathfrak m})pdpd\a$. For the rectangle billiard, the two actions should 
be all positive, so $\a\in[0,\pi/2]$ and the total measure is $\mu=2\pi {\mathfrak m} ab$. 
One obtains 
\begin{eqnarray}
\mathrm{tr}\ e^{{\hat L}t} &=&\frac{2}{\pi }
\sum_{m,n=-\infty }^{\infty }\int_{0}^{\pi/2 }e^{i(m\omega _{x}+n\omega _{y})t} d\alpha \nonumber \\
&=&\sum_{m,n=-\infty }^{\infty }J_{0}(k_{mn}l).  \label{cl-tr-rect}
\end{eqnarray}
Here $k_{mn}$ is given by Eq. (\ref{eig-rect}). This result was also
obtained by Biswas \cite{Biswas00}. The Laplace transform of the 
Bessel function $J_0(x)$ gives 
\be
g_{\text{rec}}(z) =
\sum_{m,n=-\infty }^{\infty }\frac{1}{\sqrt{z^{2}+k_{mn}^{2}}}.
\ee
The classical eigenmodes are given by $\gamma _{mn}=ik_{mn}$. These
classical eigenmodes are discrete. The above trace formula of $g_{\rm rec}(z)$
diverges as $2abk_{N}/\pi $ with $k_{N}$ the cutoff of $k_{mn}$.

\emph{Equilateral triangle billiard} For a particle in an
equilateral triangle billiard, the classical dynamics is integrable. The
phase space surface is a regular hexagon with parallel sides identified, and
thus has the topology of a torus \cite{Berry99}. The classical eigenmodes
are given by $\gamma _{mn}=ik_{mn}$ with $k_{mn}$ given by Eq. (\ref{eig-tri}%
) and $m,n=0,\pm 1,\pm 2,\cdots $. The trace of the resolvent is 
\begin{equation}
g_{\text{tri}}(z)={1\over 3}\sum_{m,n=-\infty }^{\infty }\frac{1}{\sqrt{z^{2}+k{%
_{mn}^{2}}}}.
\end{equation}
The trace of the resolvent can be derived from the following expression
\bea
\mathrm{tr}\ e^{{\hat L}t}&=&\sum_{m,n=-\infty }^{\infty }
\int_{0}^{\pi/3}
e^{i(4\pi/3a)[m\cos\a +n\cos(\pi/3-\a)]l}{d\alpha \over \pi}\nonumber \\
&=&{1\over 3}\sum_{m,n=-\infty }^{\infty }
J_0(k_{mn}l)  \label{cl-tr-tri}
\eea
Though this is similar to that for the rectangle billiard,
we are unable to give a derivation nor a physical interpretation
for this expression.

\emph{Circle billiard} For a particle in a circle
billiard of radius $a$, the Hamiltonian is $H=p_{r}^{2}/2{\mathfrak m}+p_{\theta
}^{2}/2{\mathfrak m}r^{2}$. The actions for the angular and radial freedoms are $%
I_{\theta }=p_{\theta }$ and $I_{r}=\pi ^{-1}(p^2a^{2}-I_{\theta
}^{2})^{1/2}-\pi ^{-1}I_{\theta }\arccos (I_{\theta }/pa)$ 
with $p=\sqrt{2{\mathfrak m}E}$ \cite{Jalabert}. 
We parameterize the two actions as $I_\t=pa\cos \alpha $ and
$I_r=pa(\sin\a-\a\cos\a)/\pi$ with $0\leq \a\leq \pi$
since $I_r$ is positive and $I_\t$ can take positive or negative value \cite{ODA}.
The frequencies are $\omega _{r}=\pi v/a\sin \alpha $ and 
$\omega _{\theta }=\alpha v/a\sin \alpha $. 
They are both positive as required.
Since the Jacobian determinant is
$|\partial(I_\t,I_r)/\partial(p,\a)|={1\over \pi}pa^2\sin^2\a$,
so $d\mu=4\pi a^2\d(E-p^2/2{\mathfrak m})p\sin^2\a dp d\a$ and $\mu=2\pi^2{\mathfrak m}a^2$. Thus one has 
\begin{equation}
\mathrm{tr}e^{{\hat L}t}={2\over \pi}
\sum_{m,n=-\infty }^{\infty }
\int_{0}^{\pi}
e^{i(m\alpha +n\pi )l/a\sin \alpha }\sin^2 \alpha d\alpha  \label{cl-tr-cir}
\end{equation}
and 
\begin{equation}
g_{\text{cir}}(z)=\sum_{m,n=-\infty }^{\infty }\Xi (z;m,n)
\label{resol-cir-int}
\end{equation}
with 
\begin{equation}
\Xi (z;m,n)\equiv {2\over \pi}\int_{0}^{\pi}\frac{az \sin^4\alpha d\alpha }{a^2z^2\sin^2
\alpha +(m\alpha +n\pi )^2}.
\end{equation}
Here $z=\gamma -is$ with $\gamma>0$. 

Simple expressions can be obtained for $\Xi(z;m,n)$ with $m=0$.
One has $\Xi (z;0,0)=1/az$. 
For $n\neq 0$, one has $\Xi(z;0,n)=\Xi(z/n;0,1)/n$ with
\be
\Xi (z;0,1)={2\pi+\sqrt{a^2z^2+\pi^2}\over [\pi+\sqrt{a^2z^2+\pi^2}]^2}
{az\over \sqrt{a^2z^2+\pi^2}}. \label{Xi-01}
\ee
Here we have used the following integral
\[
\int_0^{\pi}{\cos 2n\a d\a\over 1+z^2\sin^2\a}={\pi\over \sqrt{1+z^2}}
\Big({z\over 1+\sqrt{1+z^2}}\Big)^{2n}.
\label{int-10}
\]
From the above expressions of $\Xi(z;0,n)$, one gets the classical eigenmodes
\begin{equation}
\gamma _{0n}=in\pi/a,\quad n\geq 0.
\end{equation}

For $m\neq 0$, no simple expression is found for $\Xi(z;m,n)$,
though the classical eigenmodes can still be obtained 
from its integral representation.
For $m,n\neq 0$, the denominator in the integral representation of $\Xi(z;m,n)$ 
may have two roots for certain $z$ being purely imaginary and $\a\in(0,\pi)$.
If these two roots are equal, $\Xi(z;m,n)$ will diverge. 
That is, $z=\pm if(\a_0)$ with $f(\a)=(m\a+n\pi)/a\sin\a$ and its first
derivative $f'(\a)=[m-(m\a+n\pi)\cot\a]/a\sin\a$ vanishes at $\a_0$.
The classical eigenmodes are thus given by 
\begin{equation}
\gamma _{mn}=im\sqrt{1+u^{2}}/a  \label{circle-cl}
\end{equation}
with $u$ the solution of the transcendental equation 
\be
u=n\pi /m+\arctan u. 
\ee

\begin{figure}
\center{\includegraphics [angle=0,width=8cm]{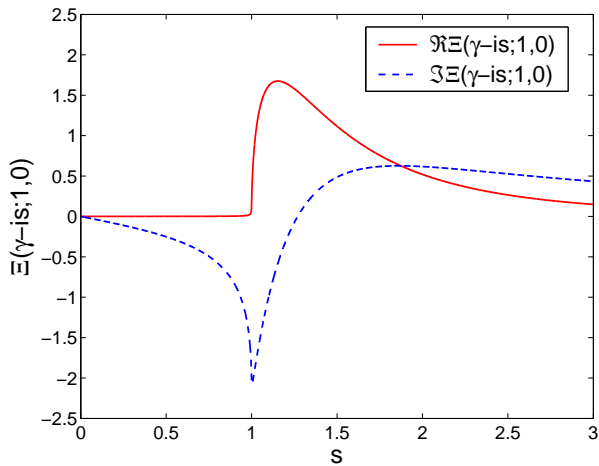}}
\caption{Function $\Xi(\c-is;1,0)$ with $\c=0.001$ and $a=1$. 
The solid line is real part $\Re\Xi(\c-is;1,0)$ and the dashed line is
the imaginary part $\Im\Xi(\c-is;1,0)$.}
\label{fig1}
\end{figure}

For $m\neq 0$ and $n=0$, 
one has $\Xi(z;m,0)=\Xi(z/m;1,0)/m$. 
$\Xi(\c-is;1,0)$ is free of divergence even when $\gamma\to 0$ and
$\Re\Xi(-is;1,0)=0$ for $s\leq 1$. 
We use $\Re$ for the real part and $\Im$ for the imaginary part throughout the paper.
For $s>1$, $\Re\Xi(-is;1,0)>0$
and has a maxmum at $s\simeq 1.1525$. 
The function $\Xi(\c-is;1,0)$ with a small $\c$ is evaluated 
numerically and plotted in Fig. \ref{fig1}.
The resonance-looking hump in $\Re\Xi(-is;1,0)$ 
 gives rise to continuous classical
eigenmodes and may be the reason for the apparent nonzero Lyapunov exponent 
\cite{Vega93} in the circle billiard and exponential short-time decay in
circle billiard with small holes \cite{Bauer90}. 
\emph{We emphasize that these classical eigenmodes bear no resemblance 
to the quantum eigenmodes.} The first few discrete classical 
eigenmodes are listed in Table I.

\begin{table}[htbp]
\caption{A few low-lying quantum and classical eigenmodes of the circle
billiard. }
\label{tab-circle}\centering                      
\begin{tabular}{c|c|c}
\hline\hline
$k_{mn}$ for NBC & $\Im \gamma _{mn}$ & $k_{mn}$ for DBC \\ \hline
0 & 0 & 2.404825558 \\ \hline
1.841183781 & 3.141592654 & 3.831705970 \\ \hline
3.054236928 & 4.603338849 & 5.135622302 \\ \hline
3.831705970 & 5.943387741 & 5.520078110 \\ \hline
4.201188941 & 6.283185307 & 6.380161896 \\ \hline
5.317553126 & 7.221364982 & 7.015586670 \\ \hline
5.331442774 & 7.789705768 & 7.588342435 \\ \hline
6.415616376 & 8.460248568 & 8.417244140 \\ \hline
6.706133194 & 9.206677698 & 8.653727913 \\ \hline
7.015586670 & 9.424777961 & 8.771483816 \\ \hline
7.501266145 & 9.671732536 & 9.761023130 \\ \hline
8.015236598 & 10.566779006 & 9.936109524 \\ \hline\hline
\end{tabular}
\end{table}

For generic systems, the frequencies $\omega
_{i}$ can not be unentangled with each other since the phase space is not a
torus. Eq. (\ref{trace-freq}) is not useful to obtain the classical
eigenmodes. However the trace of the resolvent can be expressed in terms of
POs which are the periodic solutions of the classical
equations of motion. For hyperbolic systems, Cvitanovi\'c and Eckhardt \cite
{Cvitanovic} obtained the classical trace formula for the resolvent which is
similar to the Gutzwiller semiclassical trace formula \cite{Gutzwiller}.
The classical trace formula for polygonal billiards is derived recently by 
Biswas \cite{Biswas00}. Nevertheless, exact classical trace formulas can be
derived directly from Eq. (\ref{trace-freq}) for integrable billiards.

For the \textbf{rectangle billiard}, the fluctuation part of the classical 
trace formula can be directly derived from Eq. (\ref{cl-tr-rect}) as
shown in the Appendix as 
\begin{eqnarray}
\delta g_{\text{rec}}(z) &=&\frac{2ab}{\pi }\Bigg[{4}\sum_{m,n=1}^{\infty }{%
\frac{{e^{-z{L_{mn}}}}}{{L_{mn}}}}  \nonumber \\
&&{+2}\sum_{m=1}^{\infty }\left( {\frac{{e^{-z{L_{m0}}}}}{{L_{m0}}}+\frac{{%
e^{-z{L_{0m}}}}}{{L_{0m}}}}\right) \Bigg].  \label{resol-rect}
\end{eqnarray}
Here the length of the POs is given by 
\begin{equation}
{L_{mn}=}2\sqrt{(ma)^{2}+(nb)^{2}}.  \label{Lp-rect}
\end{equation}

For the \textbf{equilateral triangle billiard}, along the same lines of derivation
for the rectangle billiard, one gets
\begin{equation}
\delta g_{\text{tri}}(z)=\frac{\sqrt{3}a^{2}}{2\pi }\left(
6\sum_{m,n=1}^{\infty }{\frac{{e^{-z{L_{mn}}}}}{{L_{mn}}}}%
+6\sum_{m=1}^{\infty }\frac{e^{-zL_{m0}}}{L_{m0}}\right)  \label{resol-tri}
\end{equation}
with 
\begin{equation}
L_{mn}=a\sqrt{3(m^{2}+n^{2}+mn)}.  \label{Lp-tri}
\end{equation}

For the \textbf{circle billiard}, the classical trace formula can be derived
directly from Eq. (\ref{cl-tr-cir}) as shown in the Appendix as
\begin{equation}
\delta g_{\text{cir}}(z)=8 a^2 \sum_{n=1}^{\infty }\sum_{m=2n}^{\infty }f_{mn}{\frac{%
{\sin^4(n\pi/m)}}{L_{mn}}e^{-z{L_{mn}}}}  \label{resol-cir}
\end{equation}
with
\bea
f_{mn}&=&2\quad {\rm for}\quad m>2n,\nonumber \\
&=&1\quad {\rm for}\quad m=2n,\nonumber \\
L_{mn}&=&2m a\sin (n\pi /m).  \label{Lp-cir}
\eea

In summary the above represents exact trace formulas for the classical dynamics of integrable billiards.

\section{Semiclassical derivation of quantum correlations}

A semiclassical consideration can lead to the understanding of the quantum
correlation. For large $k$, the semiclassical DOS for integrable 
polygonal billiards is \cite{Brack}
\begin{equation}
\delta \rho _{\epsilon }(k)\simeq \sqrt{\frac{k}{2\pi ^{3}}}%
\Re \sum_{p}a_{p}e^{(ik-\varepsilon )L_{p}-i\pi /4}/\sqrt{L_{p}}
\end{equation}
The auto-correlation $C_{\rho_{\epsilon } }(s)$ can be written as the sum of diagonal
and off-diagonal terms, $C_{\rho_{\epsilon } }(s)=C^{\epsilon }_{\text{diag}}(s)
+C^{\epsilon }_{\text{off}}(s)$.
For integrable systems, the actions are uncorrelated since there is a direct
relation between actions and the quantum spectrum. This leads to the
cancelation of the off-diagonal terms, 
\begin{equation}
C_{\rho_{\epsilon } }(s)\simeq \frac{K_{0}}{4\pi ^{3}}
\Re \sum_{p}
\frac{a_{p}^{2}}{L_{p}}e^{-(2\epsilon -is)L_{p}}.
\end{equation}
Note that the above summation is over POs with different $L_{p}$. Different
POs with the same $L_{p}$ are grouped together. For the rectangle billiard
with irrational aspect ratio, $a_{p}=4ab$ for all PO families $L_{mn}$ 
with nonzero $m$ and $n$ except for $L_{m0}$ and $L_{0n}$. Thus 
\begin{equation}
C_{\rho_{\epsilon } }(s)\simeq \frac{\left\langle a_{p}\right\rangle K_{0}}{8\pi ^{2}}
\Re \delta g_{\text{rec}}(2\epsilon -is).
\end{equation}

With normalization at $s=0$, $C_{\rho _{\epsilon }}(s)$ and $\Re \delta 
g(2\varepsilon -is)$ are almost identical with each other though
the relative heights of their peaks may be slightly different for rectangle
billiards with rational aspect ratio or the equilateral triangle billiard.
This is due to the fact that $a_{p}$ is no longer the same for all POs with
different length for these billiards. This gives rise to extra fine
structure in $C_{\rho_{\epsilon } }(s)$. We point out that the appearance of fine
structures in $C_{\rho _{\epsilon }}(s)$ is not an indication of discrete
nature of the quantum spectrum as claimed in \cite{Bogomolny,Smilansky},
rather that $a_{p}$ is not uniform.

For the circle billiard, there is no exact trace formula for $\rho (k)$ due
to the lack of analytic expression for the zeros of the Bessel functions.
For the billiard with DBC, the fluctuation part of the semiclassical DOS is 
\cite{Brack96,Brack} 
\be
\delta \rho _{\epsilon }(k)\simeq 4a^2\sqrt{\frac{k}{2\pi}}
\Re \sum_{n=1}^{\infty}\sum_{m=2n}^{\infty }
f_{mn}{\frac{{\sin^2(n\pi/m)}}{L_{mn}^{1/2}}}e^{i\Phi_{mn}}  \label{cir-DBC}
\ee
with $\Phi_{mn}=(k+i\epsilon)L_{mn}+\nu_m$, $\nu _{m}={(6m-1)\pi /4}$, 
and $f_{mn}$ given in Eq. (\ref{Lp-cir}).
This trace formula will give the WKB quantization \cite
{Jalabert,Brack} 
\begin{equation}
k_{0n}=(n+3/4)\pi
\end{equation}
which approximates the zeros of $J_{0}(z)$ and 
\begin{equation}
k_{mn}=m\sqrt{1+u^{2}}  \label{cir-WKB}
\end{equation}
with 
\[
u=(n+3/4)\pi /m+\arctan u 
\]
for the approximate zeros of $J_{m}(z)$ with $m>0$. For the billiard with
NBC, the factor $3/4$ should be replaced by $1/4$. With either boundary
condition, the diagonal part of the auto-correlation is 
\begin{equation}
C_{\rho_{\epsilon } }(s)\simeq \frac{16a^4K_{0}}{\pi }
{\Re }\sum_{n=1}^{\infty
}\sum_{m=2n}^{\infty }{\frac{{\sin^4(n\pi/m)}}{L_{mn}}e^{-(2\epsilon -is)z{%
L_{mn}}}.}  \label{q-cor-cir}
\end{equation}
For the circle billiard, one has
\be
C_{\text{diag}}^{\epsilon
}(s)\simeq \frac{4\pi a^2K_{0}}{8\pi^2 }\Re \delta g_{\text{cir}}(2\epsilon -is).
\ee
\emph{Thus the quantum correlation is determined 
by the classical eigenmodes.}

\section{Numerical calculation of quantum correlations}

\begin{figure*}[htbp]
\center{\includegraphics [angle=90,width=15cm]{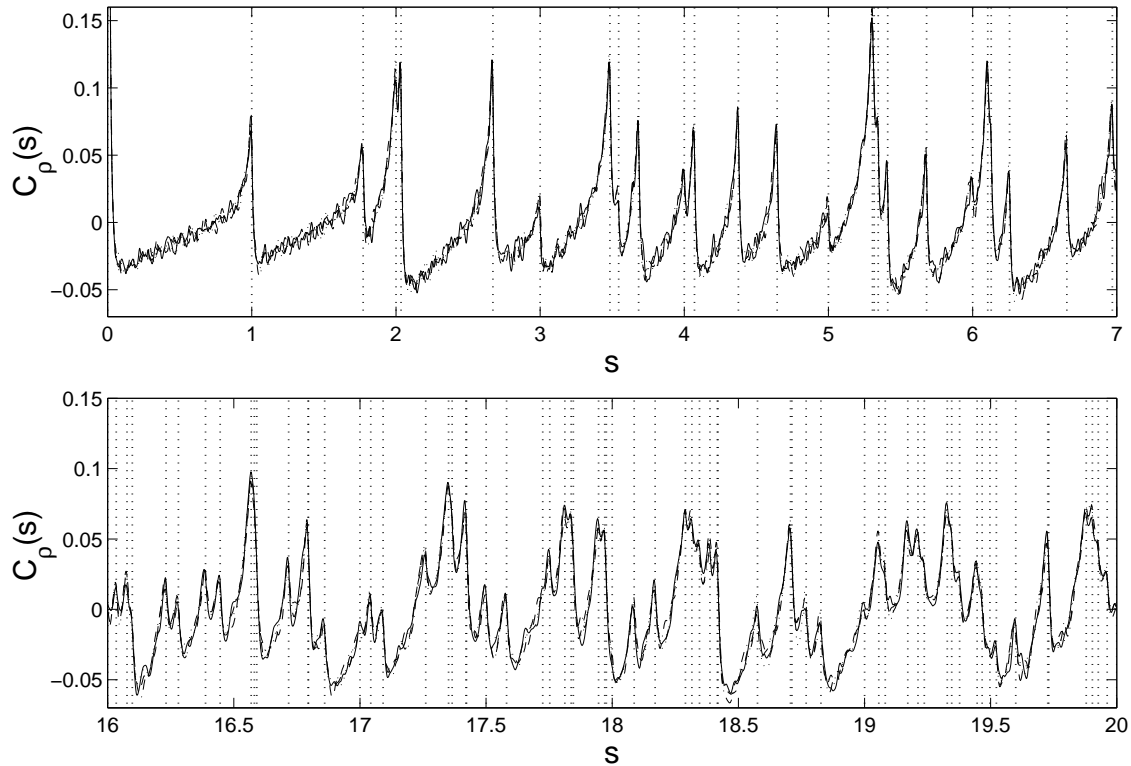}}
\caption{Auto-correlation $C_{\rho_{\epsilon } }(s)$ of the rectangular billiard with
sides $a=\pi$ and $b=\protect\sqrt{\pi }$. Here $\epsilon=0.005$. $10^{6}$
quantum eigenmodes with $0<k_{mn}<1500$ are included in $\delta\rho_{\epsilon }(k)$. The
auto-correlations $C_{\rho_{\epsilon } }(s)$ for $\delta\rho_{\epsilon }(k)$ in four equal intervals $%
[0,375]$, $[375,750]$, $[750,1125]$, $[1125,1500]$ are calculated. They are
normalized such that $C_{\rho_{\epsilon }}(0)=1$ and all collapse to the same
curve. Vertical dashed lines indicate the location of the classical
eigenmodes.}
\label{fig2}
\end{figure*}

\begin{figure*}[htbp]
\center{\includegraphics [angle=-90,width=15cm]{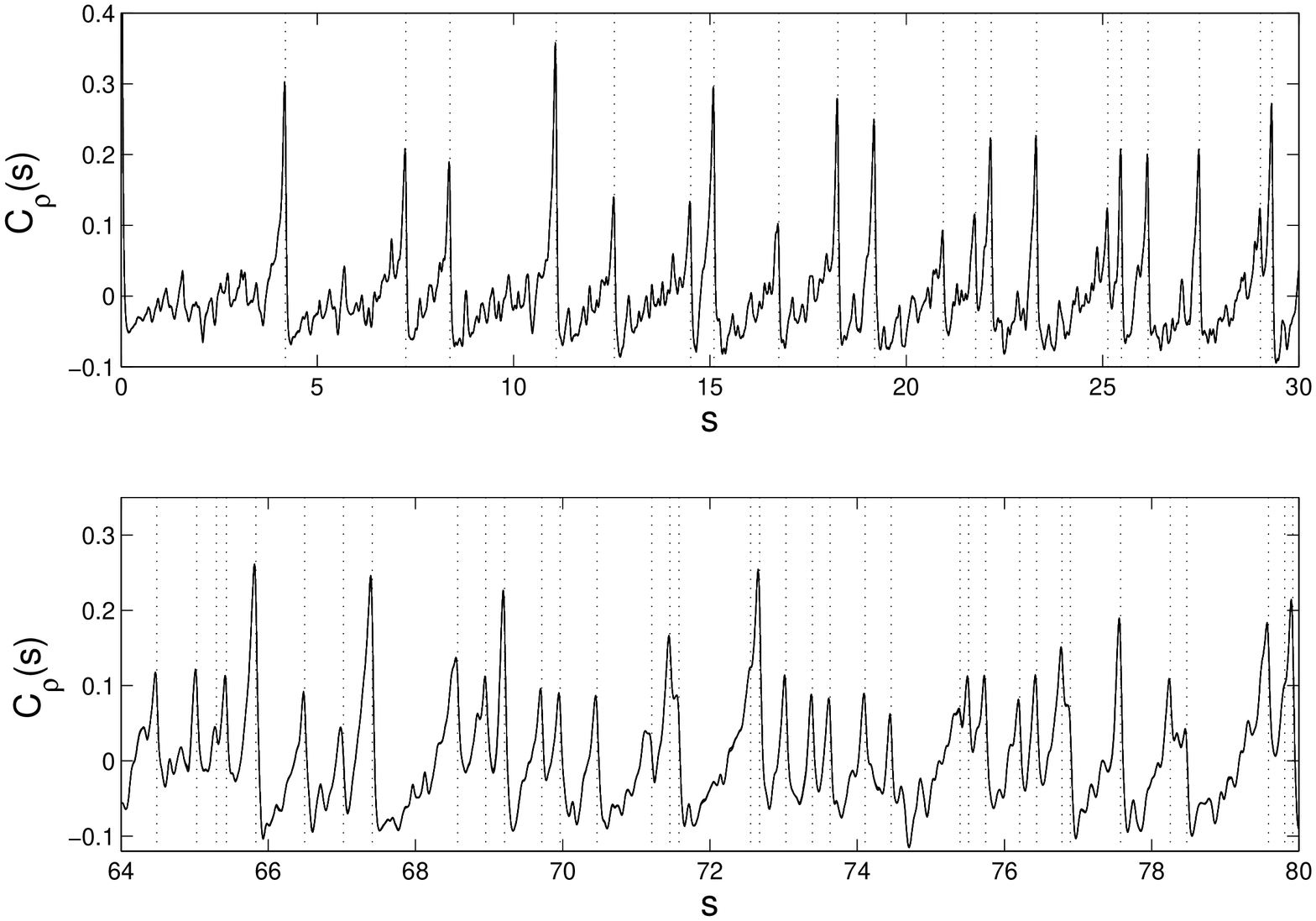}}
\caption{Auto-correlation $C_{\rho_{\epsilon } }(s)$ of the equilateral triangle
billiard with side length $a=1$. About $10^{6}$ quantum
eigenmodes with $0<k_{mn}<5390$ are included in $\delta\rho_{\epsilon }(k)$ 
with $\epsilon=0.059$. The auto-correlations $C_{\rho_{\epsilon } }(s)$ for 
$\delta\rho_{\epsilon }(k)$ 
in five equally divided intervals are calculated, normalized at the origin, 
and plotted together. Vertical dashed lines indicate the location of
the classical eigenmodes.}
\label{fig3}
\end{figure*}

\begin{figure*}[tbph]
\center{\includegraphics [angle=90,width=15cm]{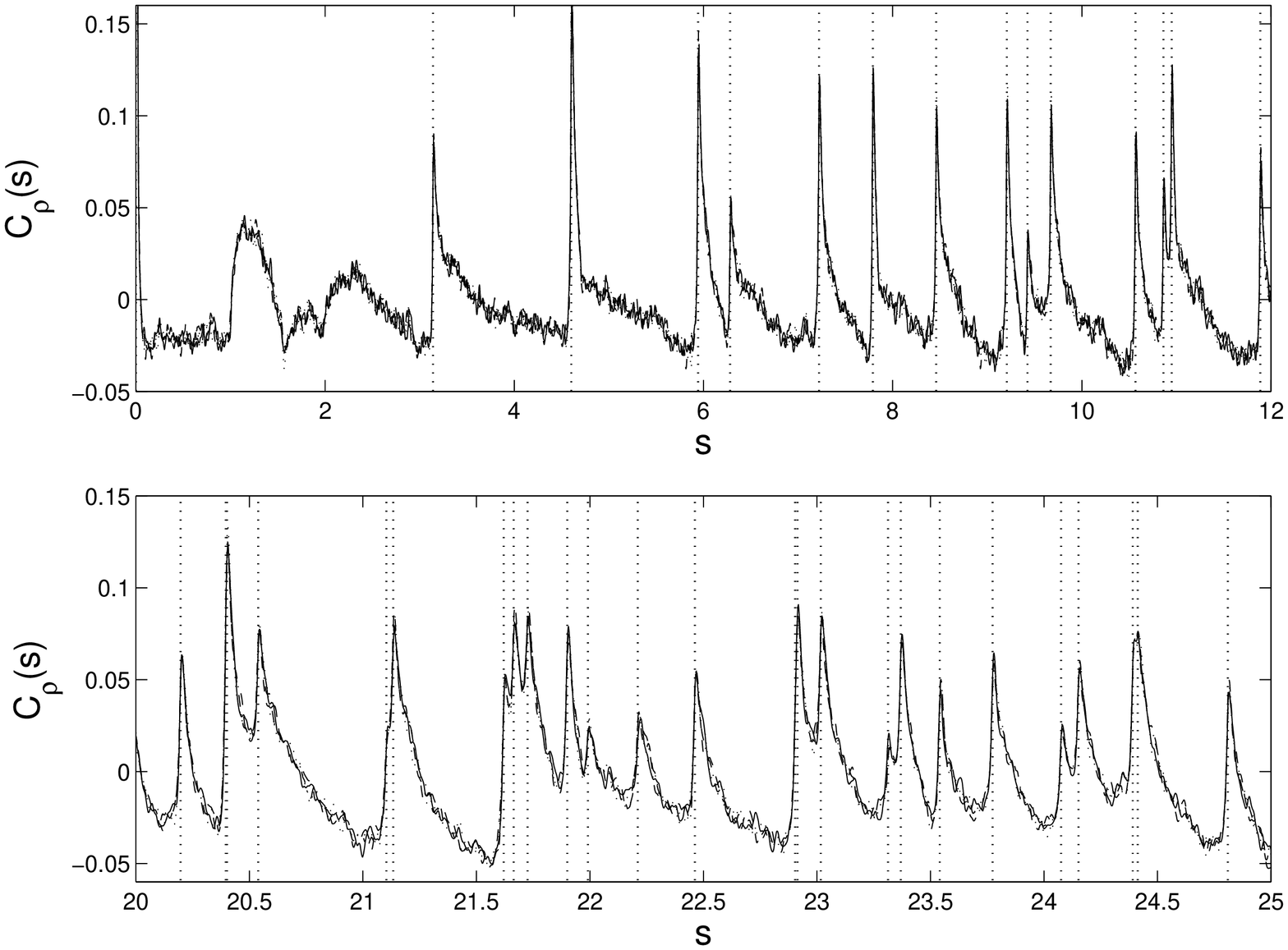}}
\caption{Auto-correlation $C_{\rho_{\epsilon } }(s)$ of the circle billiard with
radius $a=1$. About $2.3\times 10^{6}$ quantum eigenmodes with 
$0<k_{mn}<3000$ are included in $\delta \rho_{\epsilon } (k)$ with $\epsilon =0.005$. 
The auto-correlations $C_{\rho_{\epsilon } }(s)$ for $\delta \rho_{\epsilon } (k)$ in four equally 
divided intervals are calculated and normalized at the origin. 
Vertical dashed lines indicate the location of the classical
eigenmodes. The two humps in the interval $[1,3]$ correspond to $\Re\Xi(2\epsilon-is;1,0)$ 
and $\Re\Xi(2\epsilon-is;2,0)$, respectively.}
\label{fig4}
\end{figure*}

Numerical calculations of $C_{\rho _{\epsilon }}(s)$ were carried out to
confirm the results discussed here. Results for the rectangle billiard with
DBC are shown in Fig. \ref{fig2}. For the
equilateral triangle billiard of side length $a=1$ with DBC, the spectral 
correlation is shown in Fig. \ref{fig3}. Since the spectrum
of the rectangle and equilateral triangle billiards with NBC is almost the
same as the counterparts with DBC, their spectral correlations are
essentially the same. The spectral correlations for the circle billiard 
of unit radius is shown in Fig. \ref{fig4}. 
About $10^{6}$ quantum eigenvalues are used in the above calculations.

As seen from the above figures, the auto-correlations $C_{\rho_{\epsilon } }(s)$ are
almost the same for different spectra segments, confirming the 
correlation invariance.
It is quite obvious that the peaks of the quantum wave vector correlation are
located at $s=\Im \gamma _{mn}$ with $\gamma _{mn}$ the classical eigenmodes.

In Fig. \ref{fig2}, the rectangle billiard has an irrational aspect ratio. In this
case, almost all the PO families have the same summation weight. With fixed
quantum spectral width $\epsilon $, for a finite segment of quantum
spectrum, the difference between the correlation $C_{\rho _{\epsilon }}(s)$
and the classical trace $\Re \delta g(2\epsilon -is)$ can be viewed as just
noise if both are normalized at $s=0$. This noise will be reduced and
eventually disappears if either the number of the quantum eigenvalue in the
quantum spectral segment or the width $\epsilon $ increase. However for
the equilateral triangle billiard as shown in Fig. \ref{fig3}, the difference between 
$C_{\rho _{\epsilon }}(s)$ and $\Re \delta g(2\epsilon -is)$ will persist
even if the center $K_{0}$ of the quantum spectral segment is pushed to
infinity. Actually, there are five curves with different $K_{0}$ plotted in
Fig. \ref{fig3}. They are hardly distinguishable from each other. The classical trace
$\Re \delta g(2\epsilon -is)$ has no structure between classical eigenmodes.
The presence of small peaks between classical eigenmodes 
in $C_{\rho _{\epsilon }}(s)$ is what we called fine structure 
in the previous section and is due to the nonuniformity of $a_{p}$.

An interesting point about the circle billiard is that there are some
resonance-like humps in the correlation as shown in Fig. \ref{fig4} between the origin
and the first nonzero discrete classical eigenmodes $\gamma _{01}=i\pi $.
The hump between $[1,\pi /2]$ is very prominent. These humps may be the
precursor of Ruelle-Pollicott resonances in chaotic systems such as the
Bunimovich stadium.

Two comments are in order. The first is that though it is tempting to make a
one-one correspondence between the quantum and the classical eigenmodes for
integrable systems, actually this is not true. For any segment of the
quantum spectra, the auto-correlation will lead to the same low-lying classical
eigenmodes as exemplified by the invariance of the quantum spectral
correlations. The second is that in general, there is no correspondence
between the trace of the classical project operator and the quantum Neumann
spectrum. One can clearly see this for the circle billiard. For the
rectangle and equilateral triangle billiards, the fact that the quantum
NBC spectrum is also the classical eigenmodes is a coincident instead of a
rule for polygonal billiard. One can view the circle billiard as the limit
of equilateral polygon with increasing number of sides.

\section{Correspondence between classical periodic orbit length spectrum
and quantum wave vector spectrum}

In the previous sections, we established a duality between quantum 
and classical eigen spectra through quantum correlations. 
This duality was also established for hyperbolic systems \cite{Pance}. 
We believe this can also be applied to generic systems.

On the other hand, the celebrated Gutzwiller trace formula indicates a 
direct duality between the quantum and classical dynamics such that the quantum DOS
is expressed in terms of classical POs. 
As a complement to the Gutzwiller trace formula, the classical PO length spectrum
can also be expressed in terms of quantum wave vector spectrum. 
This classical POs spectrum is crucial for the calculation of classical 
eigen spectrum using the Cvitanovi\'c-Eckhardt classical trace formula \cite{Cvitanovic}.

For integrable billiards, the classical 
trace formulas we derived in Sec. III assume the form
\be
\delta g(z)=\frac{2}{\pi }\sum_{p}\frac{a_{p}}{L_{p}}e^{-zL_{p}} 
\ee
with $a_p$ the area visited by the PO family with length $L_p$. 
This formula is also
valid for generic polygonal billiard \cite{Biswas00}.
In order to obtain the classical eigenmodes, the number of PO families
defined as $N(l)=\sum_{p}a_{p}\Theta(l-L_p)$ 
should be known. Here $\Theta(x)$ is the step function. $N(l)$ has
the unit of billiard area. It was proved that the POs are dense for rational
polygonal billiards \cite{Gutkin,Boshernitzan}, $N(l)\propto l^{2}$.
Recently Biswas obtained an expression of $N(l)$ in terms of the quantum
spectrum with NBC \cite{Biswas04}. Actually, $N(l)$ can also be expressed in
terms of the quantum spectrum with DBC on the same basis. Here we give a
different derivation of this connection and reveal the domain of its
validity. Often the importance of isolated POs is neglected in this
connection. We find that they are key to determine the proliferation of
the PO families. 

Since we focus on integrable billiards, diffractive POs and 
non-periodic orbits \cite{Aurich,Zheng,Biswas00a} are absent, the
wave vector DOS of integrable polygonal billiards is 
\begin{eqnarray}
\rho (k) &=&\left\langle \rho (k)\right\rangle +c_{0}\delta (k)+\frac{k}{%
2\pi }\sum_{p}a_{p}J_{0}(kL_{p})  \nonumber \\
&&-\frac{1}{2\pi }\sum_{p}b_{p}\cos k\ell_{p}.  \label{qm-trace-poly}
\end{eqnarray}
Here $c_{0}$ is the corner term and $b_p$ is the weight for 
isolated PO with length $\ell_p$ \cite{Brack}.
For two-dimensional billiards with DBC, 
the average DOS is $\left\langle \rho (k)\right\rangle =Ak/2\pi -L/4\pi $. 
Since $\r(k)=\sum_n\d(k-k_n)$, 
multiplying both sides of Eq. (\ref{qm-trace-poly}) by $2\pi lJ_{1}(kl)/k$ and 
integrating over $k$ from 
zero to infinity, one gets 
\begin{eqnarray}
N(l) &=&\frac{\pi }{4}(\alpha -2c_{0})l^{2}+\frac{L-\beta }{2}l-A
+2\pi l\sum_{n=1}^{\infty }{1\over k_n}J_{1}(k_{n}l) \nonumber \\
&&+l\sum_{n=1}^{\infty }\sum_{p^{\prime }}{\frac{b_{p^{\prime }}}{2n}}%
J_{1}(2n\pi l/\ell_{p^{\prime }}).
\end{eqnarray}
The summation over $p^{\prime }$ is for the primitive isolated PO and 
\begin{equation}
\alpha =\sum_{p^{\prime }}\frac{b_{p^{\prime }}}{\ell_{p^{\prime }}},\quad
\beta =\sum_{p^{\prime }}b_{p^{\prime }}.
\end{equation}
In the above derivation, use has been made of the 
following identities \cite{Gradshteyn}
\begin{eqnarray*}
&&\int_0^\infty dk J_0(kL_p)J_1(kl)={1\over l}\Theta(l-L_p),\nonumber \\
&&\int_0^\infty {dk\over k} \cos(k\ell_p)J_1(kl)={1\over l}\Re \sqrt{l^2-\ell^2_p}
\nonumber \\
&&\Re \sum_{n=1}^{[x]}\sqrt{x^2-n^2}={\pi\over 4}x^2-{1\over 2}x+{x\over 2}
\sum_{n=1}^\infty {1\over n}J_1(2n\pi x).
\end{eqnarray*}
The terms $\pi\a l^2/4$ and $\b l/2$ in $N(l)$ are from the first two 
terms of the third identity.
Thus one has the average
\be
\left\langle N(l)\right\rangle =\frac{\pi }{4}(\alpha -2c_{0})l^{2}+\frac{%
L-\beta }{2}l-A. 
\ee
Hence the proliferation of PO families is determined by the isolated PO and the
corner term in the quantum trace formula (\ref{qm-trace-poly}). This reveals an
intimate connection between them.

Exact expression of $N(l)$ can be obtained for the rectangle and equilateral
triangle billiards. For the rectangle billiard with DBC, the exact trace 
formula \cite{Brack} will give 
$c_{0}=1/2$, $\alpha =2$, $\beta =L=2(a+b)$. One has exactly 
\begin{eqnarray}
N(l) &=&\frac{\pi }{4}l^{2}-ab+2\pi l\sum_{m,n=1}^{\infty }\frac{1}{k_{mn}}%
J_{1}(k_{mn}l)  \nonumber \\
&&+l\sum_{n=1}^{\infty }\frac{1}{n}\Big[aJ_{1}(n\pi l/a)+bJ_{1}(n\pi l/b)%
\Big].
\end{eqnarray}
Thus $\left\langle N(l)\right\rangle =\pi l^{2}/4-ab$. We stress that there is 
no linear term and the constant term is negative in $\left\langle N(l)\right\rangle$, 
contrary to the expression obtained by Jain \cite{Jain94}. 
As a comparison, the number of quantum states is 
\begin{eqnarray}
\mathcal{N}(k) &=&{\frac{ab}{4\pi }}k^{2}-{\frac{a+b}{2\pi }}k+\frac{1}{4}+{%
\frac{k}{2\pi }}\sum_{p}{\frac{a_{p}}{L_{p}}}J_{1}(kL_{p})  \nonumber \\
&&-{\frac{1}{2\pi }}\sum_{n=1}^{\infty }\frac{1}{n}(\sin 2nka+\sin 2nkb)
\end{eqnarray}
which is obtained from Eq. (\ref{qm-trace-poly}) through an integration.
Here the summation over $p$ is given by Eq. (\ref{resol-rect}). One has
$\left\langle {\cal N}(k)\right\rangle =abk^2/4\pi-(a+b)k/2\pi+1/4$
as expected. 

For the equilateral triangle billiard with DBC \cite{Brack}, one has
$c_{0}=2/3$, $\alpha =2$, $\beta=L=3a$. Thus 
\begin{eqnarray}
N(l) &=&\frac{\pi }{6}l^{2}-{\frac{\sqrt{3}a^{2}}{4}}+2\pi
l\sum_{n=1}^{\infty }\frac{1}{k_{n}}J_{1}(k_{n}l)  \nonumber \\
&&+\frac{3}{2}l\sum_{n=1}^{\infty }\frac{a}{n}J_{1}(4n\pi l/3a)
\end{eqnarray}
with $\left\langle N(l)\right\rangle =\pi l^{2}/6-\sqrt{3}a^2/4$. The number of quantum states is 
\begin{eqnarray}
\mathcal{N}(k) &=&{\frac{\sqrt{3}a^{2}}{16\pi }}k^{2}-{\frac{3a}{4\pi }k}+{%
\frac{1}{3}}+{\frac{k}{2\pi }}\sum_{p}{\frac{a_{p}}{L_{p}}}J_{1}(kL_{p}) 
\nonumber \\
&&-{\frac{1}{\pi }}\sum_{n=1}^{\infty }\frac{1}{n}\sin (3nka/2).
\end{eqnarray}
Here the summation over $p$ is given by Eq. (\ref{resol-tri}). 
One has $\left\langle {\cal N}(k)\right\rangle =\sqrt{3}a^2k^2/16\pi-3ak/4\pi+1/3$.

Similarly, one gets
\bea
P(l)&\equiv&\sum_{p}{a_p\over L_p}\d(l-L_p)\nonumber \\
&=&{\pi\over 2}(\a-2c_0)-Ak_NJ_1(k_Nl)/l+{L-\b\over 2l}
\nonumber \\
&&+2\pi \sum_{n=1}^NJ_0(k_nl)
+\pi \sum_{n=1}\sum_{p'}{b_{p'}\over \ell_{p'}}J_0(2n\pi l/\ell_{p'}),
\nonumber \\
\\
S(l)&\equiv&\sum_{p}a_p\d(l-L_p)\nonumber \\
&=&{\pi\over 2}(\a-2c_0)l-Ak_NJ_1(k_Nl)+{L-\b\over 2}
\nonumber \\
&&+2\pi l\sum_{n=1}^NJ_0(k_nl)
+\pi l\sum_{n=1}\sum_{p'}{b_{p'}\over \ell_{p'}}J_0(2n\pi l/\ell_{p'}).
\nonumber \\
\eea
Here $k_{N}$ is a large cutoff in the quantum spectrum. 
Exact expressions of $P(l)$ and $S(l)$ can be obtained for
the rectangle and equilateral triangle billiards.

For billiard with uniform boundary condition, $a_{p}$ in the quantum trace
formula (\ref{qm-trace-poly}) are all positive, the quantity $P(l)$ is 
just the classical quantity $(\pi /2)\mathrm{tr}\exp(-{\hat L}t)$.
The Laplace transform of $P(l)$ will give the fluctuation part of 
the classical trace $g(z)$
\bea
\delta g(z) &=&(\alpha -2c_{0})/z
-\frac{2A}{\pi }\left( \sqrt{z^{2}+k_{N}^{2}}-z\right)
\nonumber \\
&&+\sum_{n=1}^{N}\frac{4}{\sqrt{z^{2}+k_{n}^{2}}}
+\sum_{n=1}\sum_{p^{\prime }}\frac{2b_{p^{\prime }}/\ell_{p^{\prime }}}
{\sqrt{z^{2}+(2n\pi /\ell_{p^{\prime }})^{2}}}.\nonumber \\
\eea
Here we ignored the term $(L-\b)/2l$ in $P(l)$ since it vanishes for 
integrable polygonal billiards and made use of the 
Laplace transforms of Bessel functions
\begin{eqnarray*}
\int_0^\infty l^{-1}J_1(k_Nl)e^{-zl}dl&=&{1\over k_N}\Big(\sqrt{z^2+k_N^2}-z\Big),\\
\int_0^\infty J_0(k_nl)e^{-zl}dl&=&{1\over \sqrt{z^2+k_n^2}}.
\end{eqnarray*}
Thus the classical eigenmodes are $\gamma _{n}=ik_{n}$
and $\gamma _{n}=i2n\pi /\ell_{p^{\prime }}$ with $\ell_{p^{\prime }}$ the
primitive isolated PO without repetition. 

{\it The results for $N(l)$, $P(l)$, $S(l)$ establish a correspondence 
between the classical PO length spectrum and the quantum 
spectrum for integrable polygonal billiards.}

For the circle billiard, no exact expression of $N(l)$ can be obtained since 
we don't have an exact quantum trace formula. Nevertheless, 
an approximation of $N(l)$ could be derived similarly through 
the semiclassical trace formula. 

We remark that the above equations may be applicable to non-integrable billiards
only if the contribution to the quantum trace formula from diffractive POs 
and non-periodic orbits can be safely ignored. 
This is not the case for generic polygonal billiards. 
For example for the $\pi /3$ rhombus billiard,
diffractive POs are nearly as dense as the PO families.
The classical spectrum is
thus different from the quantum spectrum though they may partially overlap with
each other.

\section{Conclusions and Remarks}

In this paper, we directly obtained the classical eigenmodes of the
Liouvillian dynamics and exact classical trace formula of the resolvent
 for integrable closed systems including the rectangle,
equilateral triangle and the circle billiards, and showed that the peaks in
the quantum wave vector spectral correlation coincide with the classical
eigenmodes. We also established a correspondence between the classical PO
length spectrum and the quantum wave vector spectrum for polygonal integrable
billiards.

We have shown earlier in \emph{open chaotic systems,
specifically the n-disk billiards,} \emph{that the auto-correlation of the
resonant quantum spectrum carries the fingerprints of the classical
resonances, also known as the Ruelle-Pollicott resonances}. For the Riemann
zeros which are the quantum eigenmodes on negative constant curvature, the
spectral correlation also leads to the ``classical'' eigenmodes \cite
{Leboeuf04,Lu04}. We speculate that this also holds for chaotic and generic
billiard systems. A direct duality between classical POs spectra and quantum spectra
is also obtained for integrable billiards.

While for a quantum system, there are many ways to solve the Schr\"{o}dinger
equation, the classical spectrum is very difficult to get at. Direct
numerical simulation of tracing classical trajectories is of limited use due
to the presence of the Lyapunov exponent and computer round off error. The
trace formula in terms of classical POs provide another correspondence
between the quantum and classical dynamics. The usefulness of these trace
formula to obtain classical \cite{Cvitanovic} and quantum spectra \cite
{Gutzwiller} is marred by the divergence due to the exponential
proliferation of POs. Though the classical eigenmodes are
difficult to calculate for generic systems, our approach provides a simple
way to obtain them.

Our results can be readily tested in wave-mechanical experiments, such as
the microwave analogs of quantum billiards \cite{Sridhar}. There the
measured transmission is $T(k)=\sum_{n}c_{n}/[(k-k_{n})^{2}+\epsilon
_{n}^{2}]$. Earlier work has shown in \textit{n}-disk open systems that the
peaks of the correlation $C_{T}(s)\equiv \left\langle
T(k)T(k+s)\right\rangle _{k}$ are indeed located at $s=\Im \gamma _{n}$ \cite
{Pance}. In another words, the classical eigenmodes were obtained from the
experimentally measured quantum spectrum for this open system. Similar
experiments on closed billiards should yield results that can be compared
with those discussed in the present paper. We have examined simulations of
the experimental data by taking $c_{n}$ as random numbers in $[0,2\epsilon
/\pi ]$ and set $\epsilon _{n}=\epsilon $, the same width for all
eigenvalues as we used in $\delta \rho _{\epsilon }(k)$. We find that
spectral correlations are almost identical if $c_{n}$ is constant (as in $%
\delta \rho _{\epsilon }(k)$) or uniformly distributed in $[0,2\epsilon /\pi
]$.

Classical eigenmodes play an important role in quantum dynamics. For example
they govern the time evolution of a wave packet and are related to the
problem of decoherence \cite{Zurek}. Consider the quantity 
\[
A(t)=\sum a_{n}e^{-i\hslash k_{n}^{2}t/2{\mathfrak m} } 
\]
with ${\mathfrak m}$ the particle mass, and $a_{n}$ is significant only for $%
K_{0}-\Delta <k_{n}<K_{0}+\Delta $ with $K_{0}\gg \Delta $. This quantity is
related to the wave packet revival \cite{Robinett04}. For two-dimensional
billiards, the Heinsenberg time is $t_{H}={\mathfrak m} A/h$. For the wave packet
moving with a speed $v=\hbar K_{0}/{\mathfrak m} $, we define the Heinsenberg length $%
l_{H}=vt_{H}=(A/2\pi )K_{0}$. So one has $l_{H}\rightarrow \infty $ if $%
A\rightarrow \infty $ (open systems) or $K_{0}\rightarrow \infty $ (short
wavelength). One has $\hslash k_{n}^{2}t/2{\mathfrak m} =k_{n}l-K_{0}l/2+(A/4\pi
)(k_{n}-K_{0})^{2}(l/l_{H})$. Here $l=vt$. For $l\ll l_{H}$, the last term
can be ignored, thus $|A(t)|^{2}\simeq \Big|\sum_{n}a_{n}e^{-ik_{n}l}\Big|%
^{2}$. In the case $a_{n}=1$, one has 
\[
|A(t)|^{2}\simeq \Re \int C_{\rho _{\epsilon }}(s)e^{isl}ds\simeq \Re 
\mathrm{tr}\ e^{{\hat L}t}. 
\]
So before the Heisenberg time, $|A(t)|^{2}$ is governed directly by the
classical eigenmodes $\left\{ \gamma _{n}\right\} $ of the Liouville
operator ${\hat L}$.

Boundary conditions play an important role on the dynamics. This was
addressed recently by Biswas \cite{Biswas98}. Non-identical boundary
condition on adjacent edges of the polygonal billiards can lead not only to
quantum splitting but also to \emph{quantum annihilation} of PO
families in the quantum trace formula. Such as in the case of a rectangle
billiard with rational aspect ratio and only one side is with NBC. The most
dramatic case is a square billiard with DBC on all sides except one. In this
case, the quantum annihilation leads to a massive disappearance of PO
families in the quantum trace formula. Similar to the case of quantum
splitting, the spectral rigidity will also deviate from the Poissonian value
due to quantum annihilation.

Previous studies of the quantum properties of billiard systems were focused
on the universality of level statistics, such as nearest level spacing,
spectral rigidity, etc., and the quantum-classical correspondence was
typically addressed in terms of POs \cite{Gutzwiller,Berry140}. Here we take
a different approach by examining correspondence between the classical and
quantum spectra. In general, the classical eigenmodes are different from the
quantum eigenmodes. Except for some special systems \cite
{Biswas90,Biswas93,Biswas04}, there is no direct correspondence between the
two spectra. For systems in which all the POs have the same
Lyapunov exponent, there is a {\it self-duality} between quantum eigenvalues and
the classical eigenmodes, such as the Riemann zeros. This self-duality 
leads to the {\it resurgence} \cite{Lu04,Berry307,Bohigas01,Leboeuf01,Leboeuf04} and
{\it prophecy} \cite{Lu04} of quantum eigenmodes in the spectral correlation. For
systems such as the rectangle and equilateral triangle billiards we
considered in this paper, there is an approximate self-duality between these
two spectra, the quantum eigenmodes are also the classical eigenmodes.
Prophecy \cite{Lu04} are also observed in these systems.

A general theory connecting the quantum fluctuation and the classical
spectrum of the Perron-Frobenius operator has been developed \cite
{Agam95,Keating,Sano} for diffusive systems. Due to the difficulty of
getting the classical eigenmodes for generic systems, the focus was shifted
onto the connection with random matrix theory. Since the spectral form
factor is not self-averaging for clear systems, smoothing or ensemble
average is required for the form factor of a \emph{single} system to
approach the prediction of random matrix theory \cite{Prange}. Our previous
study confirmed that for hyperbolic open systems, the quantum correlations
are determined by the classical eigenmodes \cite{Pance}. In this paper, we
have confirmed this for integrable closed billiards. Thus one is able to
extend the so-called AAA-BK theory \cite{Agam95,Keating} to a \emph{finite}
spectrum of a single closed system.

\section*{Acknowledgments}
We thank J.V. Jos\'e for discussions.
This work is supported partially by NSF-PHY-0457002.

\appendix
\section{Derivations of classical trace formulas}
In this appendix, we derive the trace formulas Eq. (\ref{resol-rect}) 
and (\ref{resol-cir}) from 
Eq. (\ref{cl-tr-rect}) and (\ref{cl-tr-cir}), respectively.  

For the rectangle billiard,
one has explicitly
\[
\mathrm{tr}\ e^{{\hat L}t}={2\over \pi}\int_0^{\pi/2} 
\sum_{m,n=-\infty}^\infty 
e^{im(\pi/a)l\cos\alpha+in(\pi/b)l\sin\alpha}d\a.
\]
Using the identity 
\begin{equation}
\sum_{m=-\infty}^\infty e^{2\pi imx}=\sum_{m=-\infty}^\infty\delta(x-m),
\end{equation}
one gets for $l>0$ 
\be
\mathrm{tr}\ e^{{\hat L}t}=
{\frac{8ab}{\pi }}\sum_{m,n=-\infty}^\infty F^{\rm rec}_{m,n}(l)
\ee
with
\be
F^{\rm rec}_{m,n}(l)=\int_0^{\pi/2} 
\delta (l\cos\alpha -2ma)\delta (l\sin\a-2nb)d\alpha.
\ee
The above defined function $F^{\rm rec}_{m,n}(l)$ is nonzero only if both $m$ and $n$ are non-negative.
So we only need to consider $m,n\geq 0$.
Using the property of the $\d$-function in the polar coordinate system \cite{delta-1}
\bea
\d^2({\bf r}-{\bf r}_0)&\equiv& \d(x-x_0)\d(y-y_0)\nonumber \\
&=&{1\over r_0}\d(r-r_0)\d(\t-\t_0)
\eea
we get for $m,n>0$,
\bea
F^{\rm rec}_{m,n}(l)&=&{1\over L_{mn}}\d(l-L_{mn})\int_0^{\pi/2} 
\d(\a-\a_{mn})d\alpha\nonumber \\
&=&{1\over L_{mn}}\d(l-L_{mn})
\eea
with $L_{mn}=2\sqrt{m^2a^2+n^2b^2}$ and $\a_{mn}=\tan^{-1}(nb/ma)$. 
For $F^{\rm rec}_{m,0}(l)$ with $m>0$, since $\a_{m0}=0$ and 
$\int_0^\infty \d(x)dx=1/2$, one gets 
$F^{\rm rec}_{m,0}(l)=\d(l-L_{m0})/2L_{m0}$. Similarly one has
$F^{\rm rec}_{0,n}(l)=\d(l-L_{0n})/2L_{0n}$ 
and $F^{\rm rec}_{0,0}(l)=\d(l)/2l$. 
So one gets
\bea
\mathrm{tr}\ e^{{\hat L}t}&=&{\frac{2ab}{\pi}}\Bigg\{{2\over l}\d(l)
+2\sum_{m=1}^\infty \Big[{\d(l-L_{m0})\over L_{m0}}
+{\d(l-L_{0m})\over L_{0m}}\Big]\nonumber \\
&&+4\sum_{m,n=1}^\infty {\d(l-L_{mn})\over L_{mn}}\Bigg\}.
\eea
The Laplace transform of the above 
expression without the first term $\d(l)/l$ will give Eq. (\ref{resol-rect}).

For the circle billiard, the trace of the Liouvillian dynamics is
\be
\mathrm{tr}\ e^{{\hat L}t} =
8 a^2 \sum_{m,n=-\infty}^\infty F^{\rm cir}_{m,n}(l)
\ee
with
\be
F^{\rm cir}_{m,n}(l)=\int_{0}^{\pi}d\alpha\sin^2\alpha \d({l\over \sin\a}-2ma)
\d({\a l\over \sin\a}-2n\pi a).
\ee
The above defined $F^{\rm cir}_{m,n}(l)$ will be nonzero only for $m,n\geq 0$.
For $m>0$, using
\bea
\d({l\over \sin\a}-2ma)&=&{l\over 2ma\sqrt{4m^2a^2-l^2}}\Big[\d(\a-\a_m)
\nonumber \\
&&+\d(\a-\pi+\a_m)\Big]\label{delta-2}
\eea
with $\a_m=\arcsin(l/2ma)$ for $m>l/2a$, we get
\bea
F^{\rm cir}_{m,n}(l)&=&{l^3\over (2ma)^3\sqrt{4m^2a^2-l^2}}
\d({\a_m l\sin\a_m}-2n\pi a)\nonumber \\
&=&{l^3\over (2ma)^4\sqrt{4m^2a^2-l^2}}
\d(\a_m-n\pi/m)
\eea
which is valid only for $m>2n$. 
Since $\a_{m}=n\pi/m$ will give $l=L_{mn}\equiv 2ma\sin(n\pi/m)$, thus 
\be
\d(\a_m-n\pi/m)=\sqrt{4m^2a^2-L_{mn}^2}\d(l-L_{mn}),
\ee
we obtain
\be
F^{\rm cir}_{m,n}(l)
={\sin^4(n\pi/m)\over L_{mn}}\d(l-L_{mn}).
\ee
One can verify that this expression is valid for all $m\geq n>0$.
For $m=0$ and $n\geq 0$, one has $F^{\rm cir}_{0,n}(l)=0$ while
$F^{\rm cir}_{m,0}(l)=l^3\d(l)/\pi(2ma)^4$ for $m>0$.
If only the contributions from $m,n>0$ are included, one gets
\bea
\mathrm{tr}\ e^{{\hat L}t}&=&8 a^2 
\sum_{n=1}^\infty 
\sum_{m=n}^\infty
{\sin^4(n\pi/m)\over L_{mn}}\d(l-L_{mn})\nonumber \\
&=&{2\over \pi}\sum_{n=1}^\infty \sum_{m=2n}^\infty
{a_{mn}\over L_{mn}}\d(l-L_{mn}).
\eea
with
\be
a_{mn}=4\pi a^2f_{mn}\sin^4(n\pi/m)
\ee
and $f_{mn}$ given in Eq. (\ref{Lp-cir}). 
A Laplace transform will give Eq. (\ref{resol-cir}).

\end{document}